\documentstyle[12pt]{article}
\begin{document}
\leftline{{\Large{\bf $n$th discrete KP hierarchy}}}
\vspace{0.5cm}
\leftline{Andrei K. Svinin}
\vspace{0.2cm}
\leftline{{\it Institute of System Dynamics and Control Theory}}
\leftline{{\it Siberian Branch of Russian Academy of Sciences}}
\leftline{{\it P.O. Box 1233, 664033 Irkutsk, Russia}}
\leftline{{\it e-mail: svinin@icc.ru}}
\begin{abstract}
We report an infinite class of discrete hierarchies which
naturally generalize familiar discrete KP one.
\end{abstract}

\section{Introduction}

The interrelation between discrete and differential integrable
hierarchies plays crucial role in obtaining solutions to the
discrete multi-matrix models \cite{bonora1}, \cite{bonora2}, \cite{aratyn}.
At a level of KP-type differential hierarchies the discrete structure
of multi-matrix models is captured by the Darboux--B\"acklund (DB) transformations.
In turn partition functions of multi-matrix models turns out to be
$\tau$-functions of differential hierarchies and are constructed as
DB orbits of certain simple initial conditions \cite{aratyn}.
The well known discrete KP (1-Toda lattice) hierarchy \cite{ueno} together with
its reductions can be viewed as a container for a set of KP-type
differential hierarchies whose solutions are generated by
DB transformations.

This paper is designed to exhibit certain class of discrete hierarchies
which generalize discrete KP
and show the relationship with general (unconstrained)
differential KP.
This relationship yields bi-infinite sequences of differential KP equipped
with
two compatible gauge transformations. We believe that these results
would be of potential interest from the physical point of view.

\section{$n$th discrete KP}

Given the shift operator $\Lambda = (\delta_{i,j-1})_{i,j\in{\bf Z}}$
one considers the Lie algebra of pseudo-difference operators
$$
{\cal D} = \left\{
\sum_{-\infty<k\ll\infty}\ell_k\Lambda^k
\right\} = {\cal D}_{-} + {\cal D}_{+}
$$
with usual splitting into ``negative" and ``positive" parts:
\[
{\cal D}_{-} = \left\{
\sum_{-\infty<k\leq -1}\ell_k\Lambda^k
\right\}\;\; {\rm and}\;\;
{\cal D}_{+} = \left\{
\sum_{0<k\ll\infty}\ell_k\Lambda^k
\right\}.
\]
We assume that entries of bi-infinite diagonal matrices $\ell_k
\equiv (\ell_k(i))_{i\in{\bf Z}}$ may depend on ``spectral" parameter
$z$ and multi-time $t\equiv(t_1\equiv x, t_2, t_3,...)$. In what follows
$\partial\equiv \partial/\partial x$ and $\partial_p\equiv\partial/
\partial t_p$.

Let us define\footnote{where $z$ acts as component-wise multiplication.}
\begin{equation}
Q = \Lambda + a_0z^{n-1}\Lambda^{1-n} + a_1z^{2(n-1)}\Lambda^{1-2n} + ...\;\;
\in{\cal D},\;\;
n\in{\bf N}
\label{matrix}
\end{equation}
with $a_k = (a_k(i))_{i\in{\bf Z}}$ being functions on $t$ only.

{\bf Proposition 1.} {\it Lax equations of $Q$-deformations
\begin{equation}
z^{p(n-1)}\partial_p Q = [Q_{+}^{pn}, Q],\;\;
p = 1, 2,...
\label{LAX}
\end{equation}
make sense.}

{\bf Proof.} One needs to use standard simple arguments
to prove correctness
of Eqs. (\ref{LAX}). It is enough to show that $[Q_{+}^{pn}, Q] =
- [Q_{-}^{pn}, Q]$ is of the same form as l.h.s. of (\ref{LAX}). $\Box$

We will refer to (\ref{LAX}) as $n$th discrete KP hierarchy.
Let us represent $Q$ as a dressing
up of $\Lambda$ by a ``wave" operator
\[
W = I + w_1z^{n-1}\Lambda^{-n} + w_2z^{2(n-1)}\Lambda^{-2n} + w_3z^{3(n-1)}
\Lambda^{-3n} + ... \in I + {\cal D}_{-}.
\]
Then $Q$-deformations are induced by $W$-deformations
\begin{equation}
\begin{array}{c}
\displaystyle
z^{p(n-1)}\partial_p W = Q_{+}^{pn}W - W\Lambda^{pn}, \\[0.4cm]
\displaystyle
z^{p(n-1)}\partial_p (W^{-1})^T = (W^{-1})^T\Lambda^{-pn} - (Q_{+}^{pn})^T(W^{-1})^T.
\end{array}
\label{SW}
\end{equation}
Define $\chi(t, z)=(z^ie^{\xi(t, z)})_{i\in{\bf Z}}$,
$\chi^{*}(t, z)=(z^{-i}e^{-\xi(t, z)})_{i\in{\bf Z}}$
with $\xi(t, z)\equiv \sum_{p=1}^{\infty}t_pz^p$ and wave vectors
\begin{equation}
\Psi(t, z) = W\chi(t, z),\;\;
\Psi^{*}(t, z) = (W^{-1})^T\chi^{*}(t, z).
\label{wv}
\end{equation}
Discrete linear system
\begin{equation}
\begin{array}{c}
\displaystyle
Q\Psi(t, z) = z\Psi(t, z),\;\;Q^T\Psi^{*}(t, z) = z\Psi^{*}(t, z), \\[0.4cm]
\displaystyle
z^{p(n-1)}\partial_p\Psi = Q_{+}^{pn}\Psi,\;\;
z^{p(n-1)}\partial_p\Psi^{*} = - (Q_{+}^{pn})^T\Psi^{*}
\end{array}
\label{DLS1}
\end{equation}
are evident consequence of (\ref{SW}) and (\ref{wv}).
Making use of obvious relations
$z\chi = \Lambda\chi$ and
$\chi_i = \partial^{i-j}\chi_j$ with $i$ and $j$ being arbitrary integers,
we deduce
\[
\Psi_i(t, z) =
z^i(1 + w_1(i)z^{-1} + w_2(i)z^{-2} + ...)e^{\xi(t, z)}
\]
\[
= z^i(1 + w_1(i)\partial^{-1} + w_2(i)\partial^{-2} + ...)e^{\xi(t, z)} \equiv
z^i\hat{w}_i(\partial)e^{\xi(t, z)} \equiv
z^i\psi_i(t, z).
\]

What we are going to do next is to establish equivalence of
$n$th discrete KP to bi-infinite sequence of differential KP copies ``glued" together
by two compatible gauge transformations one of which can be recognized
as DB transformation mapping ${\cal Q}_i \equiv \hat{w}_i\partial
\hat{w}_i^{-1}$ to ${\cal Q}_{i+n} \equiv \hat{w}_{i+n}\partial\hat{w}_{i+n}^{-1}$.
By straightforward calculations one can prove

{\bf Proposition 2.} {\it The following three statements are equivalent

(i) The wave vector $\Psi(t, z)$ satisfies discrete linear system
\begin{equation}
Q\Psi(t, z) = z\Psi(t, z),\;\;
z^{n-1}\partial\Psi = Q_{+}^n\Psi;
\label{DLS2}
\end{equation}

(ii) The components $\psi_i$ of a vector $\psi \equiv (\psi_i =
z^{-i}\Psi_i)_{i\in{\bf Z}}$ satisfy
\begin{equation}
G_i\psi_i(t, z) = z\psi_{i+n-1}(t, z),\;\;
H_i\psi_i(t, z) = z\psi_{i+n}(t, z)
\label{system1}
\end{equation}
with $H_i\equiv \partial - \sum_{s=1}^na_0(i+s-1)$ and
\[
G_i\equiv \partial - \sum_{s=1}^{n-1}a_0(i+s-1) + a_1(i+n-1)H_{i-n}^{-1} +
a_2(i+n-1)H_{i-2n}^{-1}H_{i-n}^{-1} + ... ;
\]

(iii) For sequence of dressing operators $\hat{w}_i$ following equations
\begin{equation}
G_i\hat{w}_i = \hat{w}_{i+n-1}\partial,\;\;
H_i\hat{w}_i = \hat{w}_{i+n}\partial
\label{system2}
\end{equation}
hold.
}

Consistency condition of (\ref{DLS2}) is given by Lax equation
\begin{equation}
z^{n-1}\partial Q = [Q_{+}^n, Q]
\label{first}
\end{equation}
which in explicit form looks as
\begin{equation}
\begin{array}{c}
\partial a_k(i) = a_{k+1}(i+n) - a_{k+1}(i) \\[0.4cm]
\displaystyle
+ a_k(i)\left(
\sum_{s=1}^na_0(i+s-1) - \sum_{s=1}^na_0(i+s-(k+1)n)\right),\;\;
k\geq 0.
\end{array}
\label{motion}
\end{equation}

{\bf Remark.} One-field reductions of the systems (\ref{motion})
lead to Bogoyavlenskii lattices \cite{bogoyavlenskii}
\[
\partial r_i = r_i\left(
\sum_{s=1}^{n-1}r_{i+s} - \sum_{s=1}^{n-1}r_{i-s}\right),
\;\; r_i \equiv a_0(i)
\]
including well known Volterra lattice $\partial r_i = r_i(r_{i+1} - r_{i-1})$
in the case $n=2$.

Consistency condition of (\ref{system2}) is given by relations
\begin{equation}
G_{i+n}H_i = H_{i+n-1}G_i,\;\;
i\in{\bf Z}
\label{relations}
\end{equation}
which in fact are equivalent to (\ref{first}).

{\bf Proposition 3.} {\it By virtue of (\ref{system2}) and its consistency
condition, Lax operators ${\cal Q}_i$
are connected with each other by two invertible compatible gauge
transformations
\begin{equation}
{\cal Q}_{i+n-1} = G_i{\cal Q}_iG_i^{-1},\;\;
{\cal Q}_{i+n} = H_i{\cal Q}_iH_i^{-1}.
\label{similarity1}
\end{equation}
}

{\bf Proof.} By virtue of (\ref{system2}), we have
\[
{\cal Q}_{i+n-1} = \hat{w}_{i+n-1}\partial\hat{w}_{i+n-1}^{-1} =
(G_i\hat{w}_{i}\partial^{-1})\partial(\partial\hat{w}_{i}^{-1}G_i^{-1})
\]
\[
= G_i\hat{w}_{i}\partial\hat{w}_{i}^{-1}G_i^{-1} = G_i{\cal Q}_iG_i^{-1}.
\]
The similar arguments are applied to show second relation in
(\ref{similarity1}).
The mapping ${\cal Q}_i\rightarrow\tilde{{\cal Q}}_i = {\cal Q}_{i+n-1}$ we
denote as $s_1$, while $s_2$ stands for transformation
${\cal Q}_i\rightarrow\overline{{\cal Q}}_i = {\cal Q}_{i+n}$. As for
compatibility of $s_1$ and $s_2$, by virtue of (\ref{relations}), we have
\[
{\cal Q}_{i+2n-1} = G_{i+n}{\cal Q}_{i+n}G_{i+n}^{-1} =
G_{i+n}H_i{\cal Q}_{i}H_i^{-1}G_{i+n}^{-1}
\]
\[
= H_{i+n-1}G_{i}{\cal Q}_iG_i^{-1}H_{i+n-1}^{-1} =
H_{i+n-1}{\cal Q}_{i+n-1}H_{i+n-1}^{-1}.
\]
So we can write $s_1\circ s_2 = s_2\circ s_1$. The inverse maps
$s_1^{-1}$ and $s_2^{-1}$ are well defined by the formulas
${\cal Q}_{i-n+1} = G_{i-n+1}^{-1}{\cal Q}_iG_{i-n+1}$ and
${\cal Q}_{i-n} = H_{i-n}^{-1}{\cal Q}_iH_{i-n}$.   $\Box$

It is obvious that relation $s_1^n = s_2^{n-1}$ holds. Indeed the l.h.s. and
r.h.s. of this relation correspond to the same mapping
${\cal Q}_i\rightarrow {\cal Q}_{i+n(n-1)}$. The abelian group generated by
$s_1$ and $s_2$ we denote by symbol ${\cal G}$.


Rewrite second equation in (\ref{system1}) as
$
z^{n-1}H_i\Psi_i(t, z) = \Psi_{i+n}(t, z) = (\Lambda^n\Psi)_i.
$
>From this we derive
\[
z^{k(1-n)}(\Lambda^{kn}\Psi)_i = H_{i+(k-1)n}...H_{i+n}H_i\Psi_i,
\]
\[
z^{k(n-1)}(\Lambda^{-kn}\Psi)_i = H_{i-kn}^{-1}...H_{i-2n}^{-1}H_{i-n}^{-1}\Psi_i.
\]
These relations make connection between matrices of the form
\[
P = \sum_{k\in{\bf Z}}z^{k(1-n)}p_k(t)\Lambda^{kn}
\]
and sequences
of pseudo-differential operators $\{{\cal P}_i,\; i\in{\bf Z}\}$
mapping the upper triangular part of given matrix (including
main diagonal) into the differential parts of ${\cal P}_i$'s
and the lower triangular part of the matrix to the purely
pseudo-differential parts. More exactly, we have
$(P\Psi)_i = {\cal P}_i\Psi_i$, $(P_{-}\Psi)_i = ({\cal P}_i)_{-}\Psi_i$ and
$(P_{+}\Psi)_i = ({\cal P}_i)_{+}\Psi_i$,
where
\[
{\cal P}_i =
\sum_{k > 0}p_{-k}(i, t)H_{i-kn}^{-1}...H_{i-2n}^{-1}H_{i-n}^{-1} +
\sum_{k\geq 0}p_{k}(i, t)H_{i+(k-1)n}...H_{i+n}H_{i} = ({\cal P}_i)_{-} +
({\cal P}_i)_{+}.
\]

{\bf Proposition 4.}
{\it Equations $z^{p(n-1)}\partial_p\Psi = Q_{+}^{pn}\Psi,\;p=2, 3,...$
lead to $\partial_p\psi_i = ({\cal Q}_i^p)_{+}\psi_i,\;p=2, 3,...$.
}

{\bf Proof.} We have
\[
z^{p(1-n)}(Q^{pn}\Psi)_i = z^p\Psi_i = z^{i+p}\hat{w}_ie^{\xi(t, z)} =
z^i\hat{w}_i\partial^p e^{\xi(t, z)}
\]
\[
= z^i\hat{w}_i\partial^p\hat{w}_i^{-1}
\psi_i = z^i{\cal Q}_i^p\psi_i = {\cal Q}_i^p\Psi_i.
\]
Thus
\[
z^{p(n-1)}\partial_p\Psi_i = z^{i+p(n-1)}\partial_p\psi_i =
(Q_{+}^{pn}\Psi)_i = z^{p(n-1)}({\cal Q}_i^p)_{+}\Psi_i =
z^{i+p(n-1)}({\cal Q}_i^p)_{+}\psi_i.
\]
The latter proves proposition. $\Box$

Let us
establish equations managing $G_i$- and $H_i$-evolutions with respect to
KP flows.
Differentiating l.h.s. and r.h.s. of (\ref{system2}) by virtue
of Sato--Wilson equations $\partial_p\hat{w}_i = ({\cal Q}_i^p)_{+}\hat{w}_i
- \hat{w}_i\partial^p$ formally leads to evolution equations
\begin{equation}
\begin{array}{c}
\displaystyle
\partial_p G_{i} = ({\cal Q}_{i+n-1}^p)_{+}G_{i} -
G_{i}({\cal Q}_{i}^p)_{+}, \\[0.4cm]
\displaystyle
\partial_p H_{i} = ({\cal Q}_{i+n}^p)_{+}H_{i} -
H_{i}({\cal Q}_{i}^p)_{+}.
\end{array}
\label{leadsto}
\end{equation}

Standard arguments
can be used to show that Eqs. (\ref{leadsto}) are properly defined
individually.
Let us show that permutation relations (\ref{relations}) are invariant
under the flows given by equations (\ref{leadsto}). We have
\[
\partial_p(H_{i+n-1}G_i)
\]
\[
= \{({\cal Q}_{i+2n-1}^p)_{+}H_{i+n-1} - H_{i+n-1}({\cal Q}_{i+n-1}^p)_{+}\}G_i + H_{i+n-1}\{({\cal Q}_{i+n-1}^p)_{+}G_{i} - G_{i}({\cal Q}_{i}^p)_{+}\}
\]
\[
= ({\cal Q}_{i+2n-1}^p)_{+}H_{i+n-1}G_i -  H_{i+n-1}G_i({\cal Q}_{i}^p)_{+}
= ({\cal Q}_{i+2n-1}^p)_{+}G_{i+n}H_i -  G_{i+n}H_i({\cal Q}_{i}^p)_{+}
\]
\[
= \{({\cal Q}_{i+2n-1}^p)_{+}G_{i+n} - G_{i+n}({\cal Q}_{i+n}^p)_{+}\}H_i +
G_{i+n}\{({\cal Q}_{i+n}^p)_{+}H_{i} - H_{i}({\cal Q}_{i}^p)_{+}\}
= \partial_p(G_{i+n}H_i).
\]
Hence we proved that evolution equations (\ref{leadsto})
are consistent.

Define $\Phi_i = \Phi_i(t)$ via $H_i\Phi_i = 0$ or equivalently
through equation
\[
\partial \Phi_i = \Phi_i\sum_{s=1}^na_0(i+s-1).
\]
Takig into consideration second equation in (\ref{leadsto}),
we have
\[
\partial_p(H_i\Phi_i) = ({\cal Q}_{i+n}^p)_{+}H_i\Phi_i -
H_i({\cal Q}_{i}^p)_{+}\Phi_i + H_i\partial_p\Phi_i = 0.
\]
>From this we derive $\partial_p\Phi_i = ({\cal Q}_{i}^p)_{+}\Phi_i +
\alpha_i\Phi_i$ where $\alpha_i$'s are some constants.
Commutativity condition $\partial_p\partial_q\Phi_i =
\partial_q\partial_p\Phi_i$ leads to evolution equations
for KP eigenfunctions $\partial_p\Phi_i = ({\cal Q}_{i}^p)_{+}\Phi_i$,
i.e. $\alpha_i = 0$. Thus the relations
${\cal Q}_{i+n} = H_i{\cal Q}_iH_i^{-1}$ defines DB
transformations with eigenfunctions $\Phi_i = \tau_{i+n}/\tau_i$.
It should perhaps to recall that arbitrary eigenfunction of
Lax operator ${\cal Q}$ contains information about DB transformation
$\tau \rightarrow \overline{\tau} = \Phi\tau$ while the identity
\footnote{here conventional notations
$\{f, g\} = \partial f\cdot g - \partial g\cdot f$
and  $[z^{-1}] = (1/z, 1/(2z^2),...)$ are used.}
\[
\{\tau(t - [z^{-1}]), \overline{\tau}(t)\} +
z(\tau(t - [z^{-1}])\overline{\tau}(t) - \overline{\tau}(t - [z^{-1}])
\tau(t)) = 0
\]
holds

So, we have shown that $n$th discrete KP is equivalent to
sequence of differential KP linked with each other by
two compatible gauge transformations one of which, namely,
$s_2 : {\cal Q}_i\rightarrow {\cal Q}_{i+n}$ are nothing
but Darboux--B\"acklund transformation. The problem which
can be addressed is to describe $n$th discrete KP in the language
of bilinear identities by analogy as was done for ordinary
discrete KP \cite{adler}.

\subsection*{Acknowledgments}

Many thanks to the organizers for the invitation to participate
Fourth Conference ``Symmetry in Nonlinear Mathematical Physics".

This research has been partially supported by INTAS grant 2000-15.

\end{document}